\documentclass[prb,letterpaper,aps,floatfix,superscriptaddress,twocolumn]{revtex4-1}
\usepackage{graphicx}
\usepackage{amsmath}
\usepackage{subfigure}
\newcommand{\beq}{\begin{equation}}
\newcommand{\eeq}{\end{equation}}

\newcommand{\br}{{{\bf{r}}}}

\newcommand{\bA}{{\bf{A}}}
\newcommand{\bB}{{\bf{B}}}
\newcommand{\bE}{{\bf{E}}}

\newcommand{\bq}{{\bf{q}}}

\newcommand{\bj}{{\bf{j}}}
\newcommand{\bsigma}{{\boldsymbol \sigma}}
\newcommand{\btau}{{\boldsymbol \tau}}
\newcommand{\bnabla}{{\boldsymbol \nabla}}
\newcommand{\beqa}{\begin{eqnarray}}
\newcommand{\eeqa}{\end{eqnarray}}

\begin{document}
\title{Density response in Weyl metals}
\author{I. Panfilov}
\affiliation{Department of Physics and Astronomy, University of Waterloo, Waterloo, Ontario
N2L 3G1, Canada}
\author{A.A. Burkov}
\affiliation{Department of Physics and Astronomy, University of Waterloo, Waterloo, Ontario
N2L 3G1, Canada}
\author{D.A. Pesin}
\affiliation{Department of Physics and Astronomy, University of Utah, Salt Lake City, UT 84112, USA}
\date{\today}
\begin{abstract}
We report on a study of the density response in doped Weyl semimetals or Weyl metals in the presence of an external
magnetic field. We show that the applied field leads to a contribution to the density response, which is topological in nature
and is closely related to the phenomenon of chiral anomaly. This contribution manifests in a nonanalytic nonclassical correction
to the electronic compressibility and the plasmon frequency, proportional to the magnitude of the magnetic field.
Such a nonanalytic correction to the electronic compressibility is a smoking-gun feature of Weyl metals, which clearly distinguishes them from ordinary
ferromagnetic metals.
\end{abstract}
\maketitle
\section{Introduction}
\label{sec:1}
Weyl semimetals is a new class of Dirac materials, the interest in which has grown dramatically in the last few years, following
several theoretical proposals for their realization.~\cite{Wan11,Ran11,Burkov11,Xu11}
The recent theoretical prediction~\cite{Kane12,Fang12,Fang13} and experimental observation of the very
closely related {\em Dirac semimetals}~\cite{Cava13,Shen13,Hasan13} paves the way
for the experimental observation of Weyl semimetals in the near future.

Many of the unique physical properties of Weyl semimetals may be regarded as being distinct consequences of a common underlying
phenomenon, the chiral anomaly.~\cite{Nielsen83,Volovik,Aji12,Son12,Zyuzin12,Spivak12,Grushin12,Goswami12,Qi13}
Chiral anomaly,~\cite{Adler69,Jackiw69} first discovered in the particle physics context, manifests in anomalous nonconservation of
the particle number of particular chirality in the presence of an external electromagnetic field.
It is a purely quantum-mechanical phenomenon which plays an important role in the modern understanding of
topologically-nontrivial phases of matter in general.~\cite{Ryu12,Furusaki12,Wen13}

In the context of Weyl semimetals, an important problem is to identify experimentally measurable phenomena, which can be
attributed unambiguously to chiral anomaly.
There are several proposals in the literature, involving either anomalous magnetoresistance,~\cite{Aji12,Spivak12} coupling between collective modes,~\cite{LiuQi13} or nonlocal
transport.~\cite{Pesin13}
In this paper we focus on manifestations of chiral anomaly in density response of {\em Weyl metals}, which encompass both
Weyl semimetals and doped Weyl semimetals, in which the Fermi energy is not too far from the Weyl nodes (what ``not too far" means
precisely will be explained below).
We find that the density response of Weyl metals in an external magnetic field exhibits universal features, which are unique to Weyl metals,
and may be regarded as yet another manifestation of chiral anomaly.
In particular we find that Weyl metals, placed in an external magnetic field, possess a nonanalytic correction to the electronic compressibility
and the plasma frequency, proportional to the magnitude of the magnetic field.
We argue that this is a distinguishing smoking-gun feature of Weyl metals.

A necessary condition for the existence of a Weyl semimetal is the violation of either inversion or time reversal symmetry.
This removes the Kramers degeneracy, which would otherwise preclude band touching points between pairs of nondegenerate bands.
In this paper we will focus specifically on the case of broken time reversal symmetry, i.e. a magnetic Weyl metal.
However, some of our results are in fact universal and independent of the specific realization of Weyl semimetal.

The rest of the paper is organized as follows.
In section~\ref{sec:2} we present a general expected picture of density response in a ferromagnetic metal based
on symmetry considerations.
In section~\ref{sec:3} we consider a specific simple model of a Weyl metal, based on a magnetically doped topological insulator
heterostructure.
We demonstrate explicitly that the density response of a Weyl metal, at least at low doping levels away from the nodal Weyl semimetal, is expected to be qualitatively different from that of a regular ferromagnetic
metal, discussed in section~\ref{sec:2}.
In section~\ref{sec:4} we find the plasmon collective mode frequencies of a Weyl metal in a magnetic field, explicitly considering both the cases
of a clean and dirty Weyl metal.
We finish with a brief discussion of our results and conclusions in section~\ref{sec:5}.

\section{Density response of a ferromagnetic metal in an external magnetic field}
\label{sec:2}
We will begin with a simple symmetry-based picture of the density response of a ferromagnetic metal, placed in an external magnetic field.
For simplicity of the presentation we will focus here only on the orbital effect of the external field.
We will however discuss the effects of the Zeeman splitting separately in the discussion section.

Suppose we have a ferromagnetic metal with a uniform and time-independent magnetization
along the $z$-direction.
Let us also apply an external magnetic field in the $z$-direction $\bB = B \hat z$, where $B$ can be either positive or negative
and adopt Landau gauge $\bA = x B \hat y$. We will assume that the magnetic field can have a smooth temporal and spatial variation.

Imagine integrating out electron variables to obtain an effective action for the electromagnetic field, which encodes the electromagnetic
response of the metal. To the first order in the magnetic field, this action will have the following general form
\beqa
\label{eq:1}
&&S=\sum_{\bq, i\Omega}\left[ \left(\frac{\bq^2}{8 \pi} - \frac{e^2}{2}\Pi_0(\bq, i\Omega) \right) \varphi(\bq, i \Omega) \varphi(-\bq, - i \Omega) \right. \nonumber \\
&-&\left. \frac{e^2}{2} \Pi_1(\bq, i\Omega) \varphi(\bq, i \Omega)  q_x A_y(-\bq, - i \Omega) \right. \nonumber \\
&-& \left. \frac{e^2 B}{2} \,\, \Pi_2(\bq, i\Omega) \varphi(\bq, i\Omega) \varphi(-\bq, -i \Omega) + \ldots \right] ,
\eeqa
where $\varphi(\bq, i \Omega)$ is the scalar electromagnetic potential.
The first term in Eq.~\eqref{eq:1} describes the standard electronic polarization processes in the absence of the external magnetic field.
The second and third terms couple density and magnetic responses, which is allowed by symmetry in any magnetic material.
In particular, the second term, which may be regarded as a three-dimensional (3D) generalization of the Chern-Simons term, describes the anomalous Hall response, written in
Landau gauge. 
The last term in Eq.~\eqref{eq:1} may be thought of as a magnetic field-induced correction to the electronic compressibility of the ferromagnet.
Indeed, let us consider the static limits of the $\Pi_1(\bq, i \Omega)$ and $\Pi_2(\bq, i\Omega)$ response functions.
We have
\beq
\label{eq:2}
\sigma_{xy}^{II} = \lim_{\bq \rightarrow 0} \lim_{i \Omega \rightarrow 0} e^2 \Pi_1(\bq, i\Omega) = e \left(\frac{\partial N}{\partial B} \right)_{\mu},
\eeq
and
\beq
\label{eq:3}
\frac{\partial \kappa}{\partial B} = \lim_{\bq \rightarrow 0} \lim_{i \Omega \rightarrow 0} \Pi_2(\bq, i\Omega)= \frac{\partial^2 N}{\partial \mu \partial B} =
\frac{1}{e} \frac{\partial \sigma_{xy}^{II}}{\partial \mu}.
\eeq
Here $\sigma_{xy}^{II}$ is a thermodynamic equilibrium part of the anomalous Hall conductivity~\cite{Streda} and $\kappa$ is the electronic compressibility.
The last relation can be viewed as simply a derivative of the Streda formula for the Hall conductivity Eq.~\eqref{eq:2} with respect to the chemical potential.
Thus any ferromagnetic metal has a linear in magnetic field correction to its compressibility, which is proportional to the derivative of the equilibrium part
of its intrinsic anomalous Hall conductivity with respect to the chemical potential. This is very closely related to what has been described as ``Berry phase correction
to the electron density of states in solids" by Xiao, Shi and Niu,~\cite{Niu05} and can be argued to be present in any ferromagnetic metal, based only on symmetry considerations.
In this paper we will argue that a Weyl metal is distinguished from an ordinary ferromagnetic metal by the {\em absence} of such a linear in magnetic field correction
to the compressibility. Instead, there is a nonanalytic correction to the compressibility, proportional to the {\em magnitude} of the magnetic field and which is can be associated
with the Weyl nodes. The analytic linear in the field correction appears only when the Fermi energy is sufficiently far from the Weyl nodes.
In the following section we will demonstrate this explicitly for a simple model of a Weyl semimetal in a magnetically doped multilayer heterostructure, introduced
before by one of us.~\cite{Burkov11}
\section{Density response in a Weyl metal}
\label{sec:3}
We start from the model of a Weyl semimetal in a heterostructure, made of alternating layers of topological (TI) and ordinary (NI) insulators,
doped with a sufficient concentration of magnetic impurities to produce a ferromagnetic state.~\cite{Burkov11}
Since this model has already been described in detail in a number of publications, we will be brief with the introductory details here.

The Hamiltonian has the following form
\beq
\label{eq:4}
{\cal H}(k_z) = v_F \tau^z (\hat z \times \bsigma) \cdot \left(- i \bnabla + e \bA \right) + \hat \Delta(k_z) + b \sigma^z,
\eeq
where $\hat \Delta(k_z) = \Delta_S \tau^x + \frac{1}{2}(\Delta_D \tau^+ e^{ i k_z d} + H.c.)$ is the interlayer tunneling
operator, partially diagonalized by Fourier transform with respect to the layer index and $k_z$ is the corresponding
component of the crystal momentum, defined in the first Brillouin zone (BZ) $(-\pi/d, \pi/d)$ of the multilayer superlattice.
The $\bsigma$ Pauli matrices act on the real spin degrees of freedom while the $\btau$ ones act on the pseudospin degrees
of freedom, describing the top and bottom surfaces in the TI layers.  $b$ is the spin splitting due to magnetized impurities
and $\bA$ is the vector potential of the externally applied magnetic field.
Throughout the paper we will use units in which $\hbar = c = 1$.
Finally, we will neglect the Zeeman spin splitting due to the
field, but will consider its effects in the Discussion section.

After the canonical transformation $\sigma^{\pm} \rightarrow \tau^z \sigma^{\pm},\,\, \tau^{\pm} \rightarrow \sigma^z \tau^{\pm}$,
the Hamiltonian takes the form in which the spin and pseudospin degrees of freedom decouple
\beq
\label{eq:5}
{\cal H}(k_z) =  v_F (\hat z \times \bsigma) \cdot \left(- i \bnabla + e \bA \right) + [b + \hat \Delta(k_z)] \sigma^z.
\eeq
The tunneling operator can now be diagonalized separately from the rest of the Hamiltonian, which gives
\beq
\label{eq:6}
{\cal H}_{t}(k_z) =  v_F (\hat z \times \bsigma) \cdot \left(- i \bnabla + e \bA \right) + m_t(k_z)\sigma^z.
\eeq
Here $t = \pm$ labels the two distinct eigenvalues of the tunneling operator $t \Delta(k_z)$, where
$\Delta(k_z) = \sqrt{\Delta_S^2 + \Delta_D^2 + 2 \Delta_S \Delta_D \cos(k_z d)}$, and $m_t(k_z) = b + t \Delta(k_z)$.
The corresponding eigenvectors of the tunneling operator are given by
\beq
\label{eq:7}
|u^t_{k_z} \rangle = \frac{1}{\sqrt{2}}\left(1, t \frac{\Delta_S + \Delta_D e^{-i k_z d}}{\Delta(k_z)} \right).
\eeq
The two-component spinor $|u^t_{k_z} \rangle$ is a vector in the $\tau$-pseudospin space.
To diagonalize the remaining Hamiltonian, we orient the external magnetic field in the growth direction ($\hat z$), and pick the Landau gauge $\bA = x B \hat y$. In the present model, this allows for a full analytic solution for the Landau levels. Our results are not restricted to this special orientation of the magnetic field, however. For the continuity of presentation, we defer the discussion of effect of the orientation of the field on our results until Section~\ref{sec:5}. 
It is then easily shown~\cite{Burkov12} that the eigenstates of ${\cal H}_t(k_z)$ have
the following form
\beq
\label{eq:8}
| n, k_y, k_z, s, t \rangle = v^{s t}_{n k_z \uparrow} |n-1, k_y, \uparrow \rangle + v^{s t}_{n k_z \downarrow} | n, k_y, \downarrow \rangle,
\eeq
if $B > 0$ and the $\uparrow$ and $\downarrow$ are interchanged if $B < 0$.
Here
\beq
\label{eq:9}
\langle \br | n, k_y, \sigma \rangle = \phi_{n k_y}(\br) |\sigma \rangle,
\eeq
and $\phi_{n, k_y}(\br)$ are the Landau gauge orbital wavefunctions.
$s = \pm$ labels the electron-like and hole-like sets of Landau levels
\beq
\label{eq:10}
\epsilon_{n s t} (k_z) = s \sqrt{2 \omega_B^2 n + m_t^2(k_z)} = s \epsilon_{n t}(k_z), n \geq 1,
\eeq
and the corresponding eigenvectors $|v^{s t}_n\rangle$ are given by
\beq
\label{eq:11}
|v^{s t}_{n k_z} \rangle = \frac{1}{\sqrt{2}} \left(\sqrt{1 + s \frac{m_t(k_z)}{\epsilon_{n t}(k_z)}}, - i s \sqrt{1 - s \frac{m_t(k_z)}{\epsilon_{n t}(k_z)}} \right),
\eeq
where $|v^{s t}_{n k_z} \rangle$ is a vector in the $\sigma$-pseudospin space.

The $n = 0$ Landau level is anomalous, as it is the only Landau level that does not consist of two symmetric electron and hole-like partners.
Its energy eigenvalues are given by
\beq
\label{eq:12}
\epsilon_{0 t}(k_z) = - m_t(k_z) \textrm{sign}(B)
\eeq
and
\beq
\label{eq:13}
|v^t_0 \rangle = (0,1),
\eeq
when $B > 0$ and
\beq
\label{eq:14}
|v^t_0 \rangle = (1,0),
\eeq
when $B < 0$.
The dependence of the energy of the $n=0$ Landau levels on the sign of the magnetic field in Eq.~\eqref{eq:12} is important and will play
a significant role in what follows.
To simplify the notation we can introduce a composite index $a = (s, t)$ and a tensor product eigenvector
\beq
\label{eq:15}
|z^a_{n k_z} \rangle = |v^a_{n k_z} \rangle \otimes |u^a_{k_z} \rangle.
\eeq

We can now evaluate the density response. It is convenient to use the Landau level basis, introduced above, for this calculation
and take the magnetic field to be static and uniform from the start.
Integrating out electron variables in the presence of a fluctuating scalar potential $\varphi$, which mediates Coulomb interactions between
the electrons, we obtain the following Matsubara action for $\varphi(\bq, i\Omega)$:
\beq
\label{eq:16}
S = \frac{1}{2} \sum_{\bq, i \Omega} \left[\frac{\bq^2 }{4 \pi} - e^2 \Pi(\bq, i\Omega) \right] \varphi(\bq, i\Omega) \varphi(-\bq, -i\Omega),
\eeq
where the response function $\Pi(\bq, i\Omega)$ is given by
\beqa
\label{eq:17}
\Pi(\bq, i\Omega)&=&\frac{1}{2 \pi \ell_B^2 L_z} \sum_{n, a, a', k_z} |\langle z^a_{n k_z+q} | z^{a'}_{n k_z} \rangle |^2 \nonumber \\
&\times&\frac{n_F[\xi_{n a'}(k_z)] - n_F[\xi_{n a}(k_z + q)]}{i \Omega + \xi_{n a'}(k_z) - \xi_{n a}(k_z + q)},
\eeqa
where $\ell_B = 1/\sqrt{e |B|}$ is the magnetic length.
We have assumed here that $\bq = q \hat z$, i.e. we are interested only in the collective modes, propagating along the direction
of the magnetic field, since only in this case does one get an anomalous contribution to the density response.
In this case Landau levels with different indices $n$ are not mixed in Eq.~\eqref{eq:17}.
Furthermore, it is easy to convince oneself that only $a = a'$ contributions are important in determining collective mode
dispersions, at least for small values of $q$.
In this case, we can take $\langle z^a_{n k_z + q} | z^a_{n k_z} \rangle \approx 1$ and Eq.~\eqref{eq:17} simplifies to
\beq
\label{eq:18}
\Pi(\bq, i\Omega) = \frac{1}{2 \pi \ell_B^2 L_z} \sum_{n, s, t, k_z} \frac{n_F[\xi_{n s t}(k_z)] - n_F[\xi_{n s t}(k_z + q)]}{i \Omega + \xi_{n s t}(k_z) - \xi_{n s t}(k_z + q)}.
\eeq
It is convenient to explicitly separate contributions from the anomalous $n = 0$ Landau levels and the rest. We obtain:
\beq
\label{eq:19}
\Pi(\bq, i\Omega) = \Pi_0(\bq, i\Omega) + \Pi_1(\bq, i\Omega),
\eeq
where
\beq
\label{eq:20}
\Pi_0(\bq, i\Omega) = \frac{1}{2 \pi \ell_B^2 L_z} \sum_{t, k_z} \frac{n_F[\xi_{0 t}(k_z)] - n_F[\xi_{0 t}(k_z + q)]}{i \Omega + \xi_{0 t}(k_z) - \xi_{0 t}(k_z + q)},
\eeq
is the contribution of the two $n = 0$ Landau levels and
\beqa
\label{eq:21}
\Pi_1(\bq, i\Omega)&=&\frac{1}{2 \pi \ell_B^2 L_z} \sum_{n \geq 1, s, t, k_z} \nonumber \\
&\times&\frac{n_F[\xi_{n s t}(k_z)] - n_F[\xi_{n s t}(k_z + q)]}{i \Omega + \xi_{n s t}(k_z) - \xi_{n s t}(k_z + q)}.
\eeqa
Note that the contribution of the $n = 0$ Landau levels exist only in the presence  of the magnetic field and vanishes when $B \rightarrow 0$.
On the other hand, $\Pi_1(\bq, i\Omega)$ does not vanish when $B \rightarrow 0$, with the leading correction to the $B = 0$ result going as $B^2$.
At low magnetic fields we can then assume that the leading $B$-dependence of the density response function is contained entirely in $\Pi_0(\bq, i\Omega)$,
while $\Pi_1(\bq, i\Omega)$ gives the $B = 0$ limit of the density response function.

Let us then focus on the field-dependent part, i.e. $\Pi_0(\bq, i\Omega)$.
Expanding both the numerator and the denominator of Eq.~\eqref{eq:20} to first order in $q$, we obtain
\beqa
\label{eq:22}
\Pi_0(\bq, i\Omega)&=&\frac{1}{2 \pi \ell_B^2} \int_{-\pi/d}^{\pi/d} \frac{d k_z}{2 \pi} \delta[-m_t(k_z) \textrm{sign}(B) - \epsilon_F] \nonumber \\
&\times&\frac{\frac{d m_t(k_z)}{d k_z} \textrm{sign}(B) q}{i \Omega + \frac{d m_t(k_z)}{d k_z} \textrm{sign}(B) q}.
\eeqa
Let us restrict ourselves to the most interesting situation, when the Fermi level is not far from the Weyl nodes and crosses only the $t = -$ Landau level.
The values of $k_z$, at which the crossing occurs are given by the solutions of the equation
\beq
\label{eq:23}
\Delta(k_z) = b + \epsilon_F \textrm{sign}(B).
\eeq
The dependence on the sign of the magnetic field will thus enter into the final results only through their dependence on the
Fermi energy.
Eq.~\eqref{eq:23} has two solutions, which are given by $k_z^{\pm} = \pi/d \pm k_0$, where
\beq
\label{eq:24}
k_0 = \frac{1}{d} \arccos\left[\frac{\Delta_S^2 + \Delta_D^2 - (b + \epsilon_F \textrm{sign}(B))^2}{2 \Delta_S \Delta_D}\right].
\eeq
This solution exists as long as
\beq
\label{eq:25}
b_{c1} \leq b + \epsilon_F \textrm{sign}(B) \leq b_{c2},
\eeq
where
\beq
\label{eq:26}
b_{c1} = | \Delta_S - \Delta_D|,\,\,\, b_{c2} = \Delta_S + \Delta_D.
\eeq
Introducing the Fermi velocity in the $z$-direction, corresponding to the two crossing points above
\beqa
\label{eq:27}
\pm \tilde v_F&=&\left.\frac{d \Delta(k_z)}{d k_z} \right|_{k_z = k_z^{\pm}} = \pm \frac{d}{2 [b + \epsilon_F \textrm{sign}(B)]} \nonumber \\
&\times&\sqrt{[(b + \epsilon_F \textrm{sign}(B))^2 - b_{c1}^2] [b_{c2}^2 - (b + \epsilon_F \textrm{sign}(B))^2]}, \nonumber \\
\eeqa
we finally obtain
\beq
\label{eq:28}
\Pi_0(\bq, i\Omega) = - \frac{1}{2 \pi^2 \tilde v_F \ell_B^2} \frac{\tilde v_F^2 q^2}{(i \Omega)^2 - \tilde v_F^2 q^2}.
\eeq
The dependence on the magnetic field enters through $\tilde v_F$, which depends on $\rm{sign}(B)$, and through the
magnetic length, which depends on $|B|$.

Let us note the following very important property of Eq.~\eqref{eq:28}.
Suppose the Fermi level is located not too far from the Weyl nodes, namely the following conditions are satisfied
\beq
\label{eq;29}
b_{c1} \ll b + \epsilon_F \textrm{sign}(B) \ll b_{c2}.
\eeq
In this limit, the Fermi velocity becomes
\beq
\label{eq:30}
\tilde v_F \approx \frac{d}{2} (\Delta_S + \Delta_D),
\eeq
i.e. the dependence on the Fermi energy, and, therefore, on the sign of the magnetic field, drops out.
Physically this means that the Fermi level is close enough to the Weyl nodes, so that the spectrum is
to a good approximation linear and thus Fermi velocity is independent of the Fermi energy.
In this limit, the dependence of the density response function on the magnetic field is thus nonanalytic ($\propto |B|$)
and is the same as when the Fermi level coincides with the Weyl nodes, i.e. $\epsilon_F = 0$.
It is not surprising that the $B$-dependence is nonanalytic at $\epsilon_F=0$. 
However, the fact that the nonanalytic ($\propto |B|$) correction to the electronic compressibility dominates the analytic one  
for a finite range of $\epsilon_F$ near the location of the nodes, is surprising and unexpected.
This smoking-gun feature is unique to doped Weyl semimetals, and can be used for their experimental
identification. This is one of the main results of this paper.

Let us now explore in more detail the connection between the above result for the density response, and the anomalous Hall
conductivity of a doped Weyl semimetal, more specifically its non-Fermi-surface part $\sigma^{II}_{xy}$.
An explicit expression for $\sigma^{II}_{xy}$ was derived by one of us in Ref.~\onlinecite{Burkov14}:
\beqa
\label{eq:31}
\sigma_{xy}^{II}&=&\frac{e^2}{8 \pi^2} \sum_t \int_{-\pi/d}^{\pi/d} d k_z \,\, \textrm{sign} [m_t(k_z)] \nonumber \\
&\times&\left[\Theta(\epsilon_F + |m_t(k_z)|) - \Theta(\epsilon_F - |m_t(k_z)|)\right].
\eeqa
Differentiating this expression with respect to the Fermi energy and assuming $\epsilon_F > 0$ for simplicity [$\sigma^{II}_{xy}(\epsilon_F)$ is an even function of
$\epsilon_F$], we obtain
\beq
\label{eq:32}
\frac{\partial \sigma^{II}_{xy}}{\partial \epsilon_F} = - \frac{e^2}{8 \pi^2} \sum_t \int_{-\pi/d}^{\pi/d} d k_z \textrm{sign}[m_t(k_z)] \delta(\epsilon_F - |m_t(k_z)).
\eeq
Just as before, let us assume $\epsilon_F$ is not far from the Weyl nodes.
In this case, as above, only the $t = -$ Landau level contributes to the integral in Eq.~\eqref{eq:32}.
To evaluate the integral we need to solve the equation
\beq
\label{eq:33}
|b - \Delta(k_z)| = \epsilon_F.
\eeq
This has two pairs of solutions, corresponding to $\Delta(k_z) = b \pm \epsilon_F$.
Thus we obtain
\beqa
\label{eq:34}
\frac{\partial \sigma_{xy}^{II}}{\partial \epsilon_F}&=&- \frac{e^2}{8 \pi^2} \int_{-\pi/d}^{\pi/d} d k_z \nonumber \\
&\times&\left[\delta(\Delta(k_z) - b + \epsilon_F) - \delta(\Delta(k_z) - b -\epsilon_F) \right] \nonumber \\
&=&\frac{e^2}{4 \pi^2}\left(1/\tilde v_{F+} - 1/\tilde v_{F-} \right),
\eeqa
where
\beq
\label{eq:35}
\tilde v_{F\pm} = \frac{d}{2 (b \pm \epsilon_F)} \sqrt{[(b \pm \epsilon_F)^2 - b_{c1}^2] [b_{c2}^2 - (b \pm \epsilon_F)^2]},
\eeq
coincide with the Fermi velocities, corresponding to the positive or negative $\textrm{sign}(B)$ in Eq.~\eqref{eq:27} (note that the Hall
conductivity itself is evaluated at zero field).

Now let us go back to Eq.~\eqref{eq:28} for the density response function. Let us consider its static limit, i.e. take the limits $\Omega \rightarrow 0$
and $q \rightarrow 0$ in such a way that $\Omega/ v_F q \rightarrow 0$.
In this case we obtain
\beq
\label{eq:36}
\kappa = \kappa_0 + \Pi_0(0,0) = \kappa_0 + \frac{1}{2 \pi^2 \tilde v_F \ell_B^2} = \kappa_0 + \frac{e |B|}{2 \pi^2 \tilde v_F},
\eeq
where $\kappa_0 \propto \epsilon_F^2$ is the electronic compressibility in the absence of the magnetic field.
We can now separate the compressibility into two parts, symmetric and antisymmetric with respect to changing the sign of the
magnetic field.
We obtain
\beq
\label{eq:37}
\kappa_{s} = \kappa_0 + \frac{e |B| }{4 \pi^2}(1/\tilde v_{F+} + 1/\tilde v_{F -}),
\eeq
while
\beq
\label{eq:38}
\kappa_{a} = \frac{e B}{4 \pi^2}(1/\tilde v_{F+} - 1/\tilde v_{F -}).
\eeq
An important difference between $\kappa_s$ and $\kappa_a$ is that while $\kappa_a$ is an analytic function of $B$,
$\kappa_s$ is not.
Comparing Eqs.~\eqref{eq:34} and \eqref{eq:38}, we obtain
\beq
\label{eq:39}
\frac{\partial \kappa_a}{\partial B} = \frac{1}{e} \frac{\partial \sigma^{II}_{xy}}{\partial \mu},
\eeq
which coincides with Eq.~\eqref{eq:3}.
Thus the magnetic field derivative of the analytic {\em antisymmetric} part of the electronic compressibility is equal
to the derivative of the non-Fermi-surface part of the anomalous Hall conductivity with respect to the chemical potential.
Note that $\partial \kappa_a/\partial B$ vanishes exactly when $\epsilon_F = 0$, and approximately as long as the Fermi velocity
is a constant, i.e. as long as the dispersion is linear to a good approximation.

\begin{figure}[t]
\subfigure[]{
   \label{fig:1a}
  \includegraphics[width=8cm]{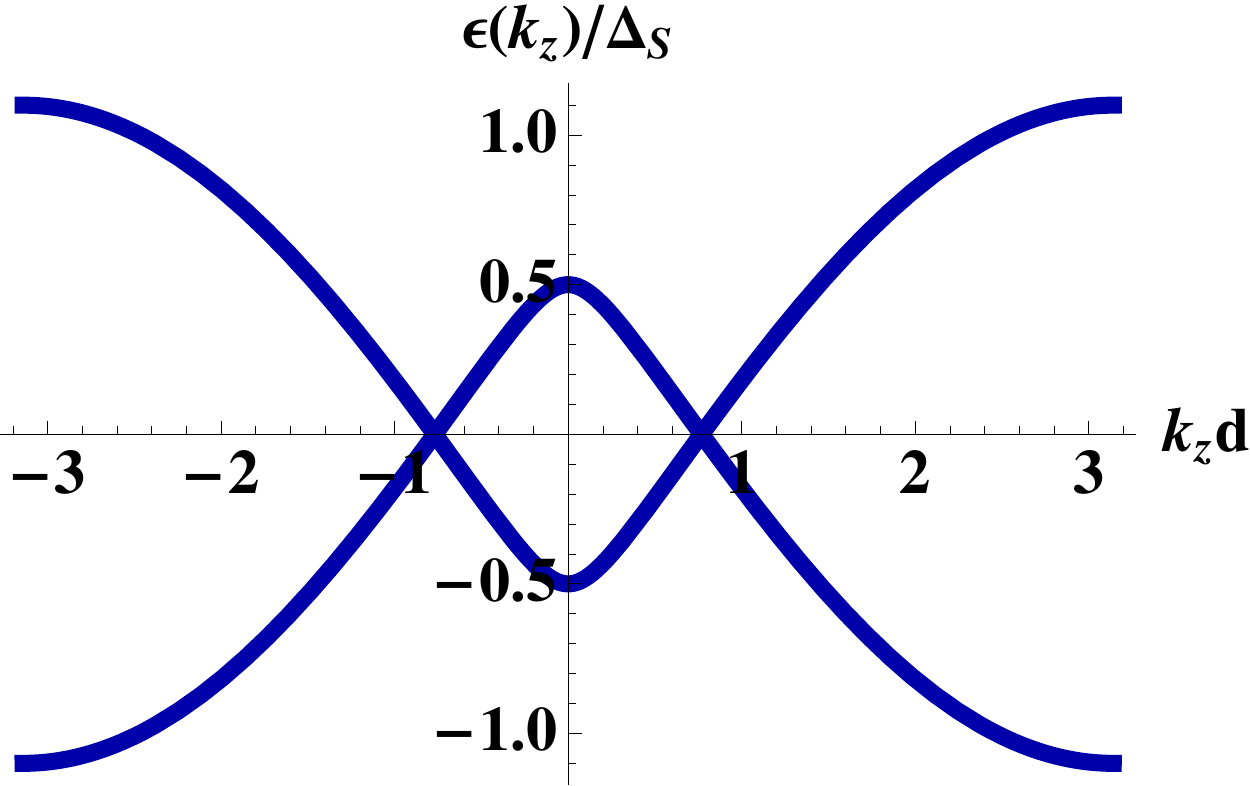}}
\subfigure[]{
  \label{fig:1b}
   \includegraphics[width=8cm]{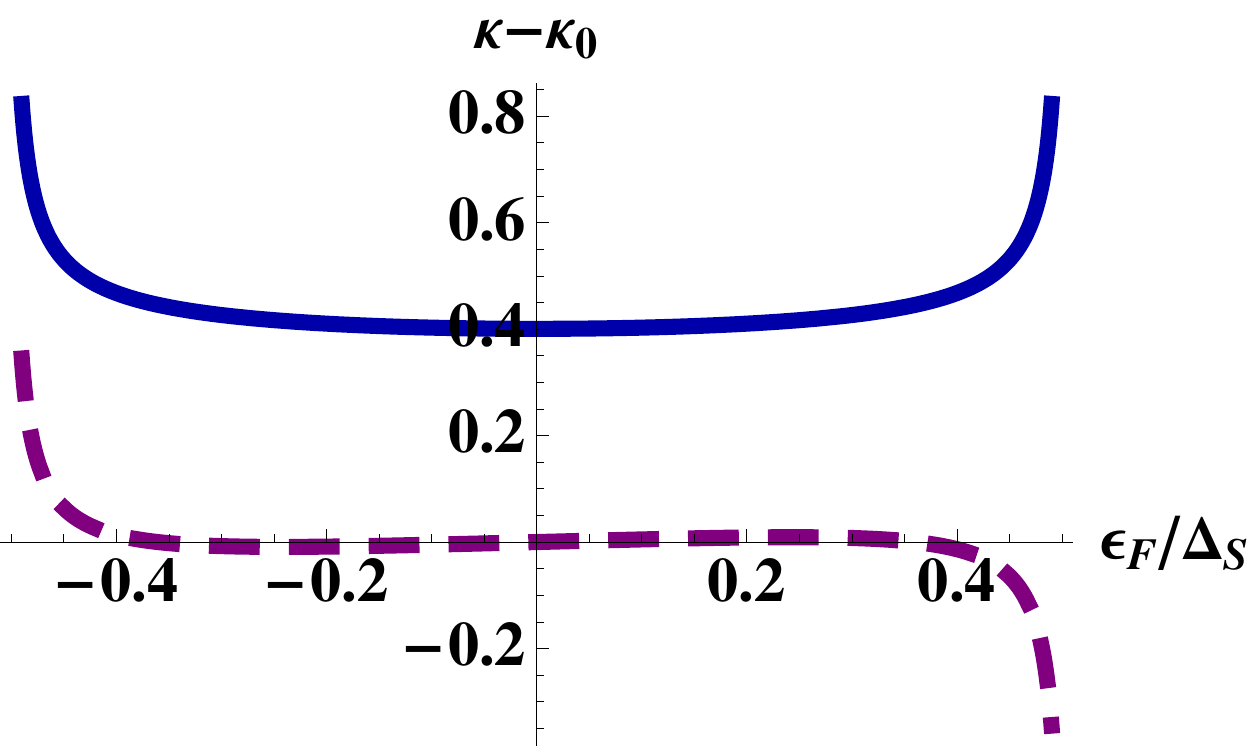}}
  \caption{(Color online). (a) Plot of the band edges along the $z$-direction in momentum space for the two bands that touch at the Weyl nodes in the absence of the
  magnetic field.  (b) Symmetric (solid line) and antisymmetric (dashed line) parts of the correction to compressibility, $\kappa - \kappa_0$, in units of $1/2 \pi \ell_B^2 \Delta_S d$.
  The antisymmetric part is negligibly small compared to the symmetric part, while the symmetric part is only weakly dependent on the Fermi energy. The apparent divergences at
  $\epsilon_F/\Delta_S = \pm 0.5$ correspond to band-edge van Hove singularities.}
    \label{fig:1}
\end{figure}

The nonanalytic symmetric part of the magnetic field dependence of the compressibility, can in turn be expressed in terms of the derivative of
$\sigma^{II}_{xy}$ with respect to the magnetization $b$, as
\beq
\label{eq:39.1}
\frac{\partial \kappa_s}{\partial |B|} = \frac{1}{e} \frac{\partial \sigma_{xy}^{II}}{\partial b},
\eeq
which is also obtained directly from Eq.~\eqref{eq:31}.
The above results are summarized in Fig.~\ref{fig:1}. 

\section{Plasmons in a Weyl metal}
\label{sec:4}
\subsection{Plasmons in a clean Weyl metal}
Let us now find the plasmon mode frequency as a function of the magnetic field.
We will first consider the case of a clean Weyl metal, ignoring impurity scattering.
The $B$-independent part of the density response function, i.e. the $B \rightarrow 0$ limit of $\Pi_1(\bq, i \Omega)$,
may be easily evaluated analytically when $\epsilon_F$ is small. In the dynamical limit $|\Omega| \gg v_F q$ one obtains
\beq
\label{eq:40}
\Pi_1(\bq, i \Omega) = \frac{\epsilon_F^2}{3 \pi^2 v_F^2 \bar v_F} \frac{\bar v_F^2 q^2}{(i \Omega)^2} = \frac{2}{3} \kappa_0 \frac{\bar v_F^2 q^2}{(i \Omega)^2},
\eeq
where $\bar v_F = \tilde v_F(\epsilon_F = 0)$. The quadratic dependence of $\Pi_1(\bq, i\Omega)$ on the Fermi energy reflects
the energy dependence of the density of states near the Weyl nodes $g(\epsilon) \propto \epsilon^2$.
Collective mode frequency is given by the solution of
\beq
\label{eq:41}
\frac{\bq^2}{4 \pi} - e^2 \Pi_0(\bq, i \Omega \rightarrow \omega + i \eta) - e^2 \Pi_1(\bq, i \Omega \rightarrow \omega + i \eta) = 0.
\eeq
Solving Eq.~\eqref{eq:41} we obtain the following expression for the plasmon frequency of a Weyl metal
\beq
\label{eq:frequency}
\omega^2 = \frac{4 e^2 \bar v_F \epsilon_F^2}{3 \pi v_F^2} + \frac{2 e^3 \tilde v_F |B|}{\pi}.
\eeq
The first term is the square of the plasmon frequency of an ordinary metal with the density of states $g(\epsilon) \propto \epsilon^2$,
while the second term contains the leading correction due to the magnetic field. Note that for small $\epsilon_F\to 0$, Eq.~(\ref{eq:frequency}) is valid for $e^2/v_F\ll 1$ only. Otherwise, the dielectric screening by higher Landau levels should in general be taken into account. 

Just as in our discussion of the electronic compressibility before, the field-dependent contribution can be separated into a nonanalytic part,
proportional to $|B|$, and an analytic part, proportional to $B$ as
\beq
\label{eq:42}
\omega_{s}^2 =  \frac{4 e^2 \bar v_F \epsilon_F^2}{3 \pi v_F^2} + \frac{e^3 |B|}{\pi}(\tilde v_{F+} + \tilde v_{F-}),
\eeq
and
\beq
\label{eq:43}
\omega_a^2 = \frac{e^3 B}{\pi}(\tilde v_{F+} - \tilde v_{F-}).
\eeq
The antisymmetric part vanishes as long as the spectrum is linear.
In this case the plasmon frequency of a Weyl metal contains a nonanalytic correction, proportional to the magnitude, but not the sign, of the applied field.
As mentioned above, this should be regarded as a general property of a Weyl metal and its smoking-gun experimental signature.

\subsection{The absence of ``hydrodynamic'' plasmon modes at low frequencies}
In this subsection we will discuss the situation when there is a considerable impurity or electron-electron scattering present near each Weyl node, while the inter-node scattering 
may still be regarded as weak. In this case can one still think of well-defined nodes and safely regard the material a Weyl metal.~\cite{Arovas13}
Specifically, we would like to see if there is a new type of collective mode with the oscillation frequency low compared to the intra-nodal momentum scattering rate, yet high compared to the typical inter-node scattering rate. We will show that if the Fermi level is such that it crosses multiple Landau levels, as can generically be expected at weak fields in a Weyl 
metal, the parameter regime for the existence of such a mode is essentially absent.
 
For simplicity, we will abandon our microscopic model of a Weyl metal here and consider a generic low-energy model of a Weyl metal with two nodes and two corresponding ``valleys" (we will refer to them by the $\pm$ chirality of the node), when the Fermi energy is away from the nodes.
We will also use a semiclassical, instead of a fully microscopic approach.

We consider a plasmon mode, propagating along an external magnetic field, $\bB$.
The fluctuations of particle density in the $\pm$ valleys in such a density wave are connected with the current fluctuation via the continuity equation,
which however contains an anomalous nonvanishing total divergence due to chiral anomaly. Neglecting the intervalley scattering for a moment, the continuity equation reads
\beq
\label{eq:44}
\frac{\partial \rho_{\pm}}{ \partial t} + \bnabla \cdot \bj_{\pm} = \pm \frac{e^3}{4\pi^2} \bE \cdot \bB.
\eeq
The equation for the current, flowing along the magnetic field, including the usual Drude contribution, but neglecting the diffusion current as we intend to consider the $\bq \to 0$
limit, is given by
\beq
\label{eq:45}
\bj_{\pm} = \sigma_D \bE \pm \frac{e^2}{4\pi^2} \mu_{\pm} \bB,
\eeq
where $\mu_{\pm}$ are the chemical potentials in the two valleys. An implicit assumption in Eq.~\eqref{eq:45} is that the intravalley impurity scattering is strong enough, so that
each valley may be assumed to be in a quasi-equilibrium state at any given moment, characterized by the corresponding chemical potential $\mu_{\pm}$.
The Drude conductivity, $\sigma_D$ is assumed to be isotropic for simplicity. The electric field entering Eq.~\eqref{eq:45} is produced by the total fluctuation in the local density
\beq
\label{eq:46}
\bnabla \cdot \bE = 4\pi (\rho_+ + \rho_-).
\eeq
Finally, the fluctuations in the density in each valley are related to the fluctuations in the chemical potentials via the usual Thomas-Fermi relations
\beq
\label{eq:47}
\rho_{\pm} = e g_B (\mu_\pm-\epsilon_{F}),
\eeq
where $\epsilon_{F}$ is the value of the unperturbed Fermi level, and $g_B$ is the density of states at the unperturbed Fermi level (which depends on the magnetic field, hence the subscript).
The magnitude of the Drude conductivity, $\sigma_D$, and the density of states, entering Eqs.~\eqref{eq:45} and ~\eqref{eq:47}, depend on the magnitude of the magnetic field, and are determined by how many Landau levels are occupied for a given $B$.
Introducing the Fermi velocity in a valley, $v_F$, evaluated at the unperturbed value of the chemical potential in that valley, we can distinguish two simple limits for $\sigma_D$ and $g_B$: That of a weak magnetic field, $\epsilon_{F} \gg v_F/\ell_B$, and strong magnetic field $\epsilon_{F} < v_F / \ell_B$.
In the former case, $g_{B} = \epsilon_F^2/ 2 \pi v_F^3$, and $\sigma_D \approx e^2 \epsilon_F^2 \tau_{\rm tr}/ 3 v_F$ ($\tau_{tr}$ being the transport scattering time), while in the latter $g_B = 1/ 4 \pi^2 v_F\ell_B^2$, $\sigma_D \to 0$.

Solving equations~\eqref{eq:44}, \eqref{eq:45}, \eqref{eq:46}, and \eqref{eq:47} simultaneously, we obtain the following equation for the plasmon frequency at zero wave vector
\beq
\label{eq:48}
\omega^2+8\pi i \sigma_D \omega - \frac{8\pi B^2}{g_B}\left(\frac{e^2}{4\pi^2}\right)^2  = 0.
\eeq
For $B=0$, this equation has a purely imaginary solution $\omega = -8\pi i \sigma_D$ which corresponds to the Maxwell relaxation of charge fluctuations with a decrement determined by the total Drude conductivity in the two valleys, $2\sigma_D$. Solutions with a non-zero real part, corresponding to either overdamped or underdamped oscillations, appear for
\beq
\label{eq:49}
 \frac{8 \pi B^2}{g_B}\left(\frac{e^2}{4\pi^2}\right)^2 > (4\pi \sigma_D)^2.
\eeq
For large magnetic fields,  such that $v_F/\ell_B \gtrsim \epsilon_{F}$, the inequality in Eq.~\eqref{eq:49} is clearly satisfied, and for
$v_F/\ell_B \gg \epsilon_{F}$ we recover the result for the plasmon frequency, obtained in the previous subsection
\beq
\label{eq:50}
\omega^2 = \frac{2e^2 v_F}{\pi \ell_B^2}.
\eeq
In the opposite limit $v_F/\ell_B \lesssim \epsilon_{F}$, it is easy to see that Eq.~\eqref{eq:49} prohibits the existence of any propagating modes. Indeed, in the limit of weak magnetic fields, the expression for $\sigma_D$ is $\sigma_D\approx e^2 \epsilon_F^2 \tau_{\rm tr}/3 v_F$, and it is easy to check that
Eq.~\eqref{eq:49} yields a condition
\beq
\label{eq:51}
 \frac{v_F}{\ell_B} > \epsilon_{F} \sqrt{\epsilon_{F} \tau_{tr}}.
\eeq
Such condition can be reconciled with the weak field one only for $\epsilon_{F} \tau_{tr}\ll 1$, which is hard to imagine to hold. Therefore, in the regime of weak magnetic fields only the usual plasmon mode exists, whose frequency is given by Eq.~(\ref{eq:frequency}).
\section{Discussion and Conclusions}
\label{sec:5}
We have so far restricted ourselves to considering only the orbital effect of the applied magnetic field.
In reality, Zeeman effect is also present and can be expected to influence the results.
Let us demonstrate that our conclusions are in fact unchanged even when the Zeeman effect is taken into account. 
Zeeman splitting adds a term $g \mu_B  B$ to the magnetization-induced spin splitting $b$.
Substituting this correction into Eq.~\eqref{eq:27} for the Fermi velocity as a function of $b$ and $\epsilon_F$, it is clear
that the effect of this correction is equivalent to a shift of the Fermi energy by $g \mu_B |B|$.
This has no effect on our results, as long as the Fermi velocity may be regarded as being independent from the Fermi energy.
Thus our conclusions regarding the absence of an analytic  linear-in-$B$ correction to the electronic compressibility of a Weyl metal
are unchanged by adding the Zeeman splitting.

Another issue that has so far not been discussed is the dependence of our results on the orientation of the magnetic field relative
to the magnetization.
On symmetry grounds we expect that the linear in magnetic field correction to the compressibility, which is common in normal metallic ferromagnets,
but is absent in Weyl metals, should be proportional to $\bB \cdot \hat m$, where $\hat m$ is the unit vector in the direction of the spontaneous 
magnetization (ignoring any intrinsic anisotropy that the material may have in the absence of the magnetization). 
The nonanalytic correction, proportional to $|B|$, on the other hand, should depend on the direction of the magnetic field less strongly,
only to the extent required by possible anisotropy of the band dispersion near the Weyl nodes.

The expected sensitivity of the proposed effect to temperature and disorder, which will ultimately determine its experimental observability, 
is clearly a nonuniversal quantitative question, the answer to which depends on the specifics of the electronic structure of a given material. 
However, one can generally expect the most important energy scale in this regard to be the exchange spin splitting $b$. Thus, as long as the temperature 
is significantly less than the critical temperature of the ferromagnet-paramagnet phase transition, and as long as the disorder-induced broadening of the 
density of states is significantly less than $b$, the anomalous correction to the compressibility should be observable.   
These conditions are easily satisfied in any material with a reasonably strong magnetic order (i.e. a critical temperature that is not too low). 

The model of a Weyl metal we have explicitly considered in this paper relies on broken time reversal symmetry to realize it.
The second class of Weyl metals is realized when inversion symmetry is broken instead.~\cite{Balents12}
In this realization of a Weyl metal, anomalous Hall effect and linear in magnetic field correction to compressibility are absent by symmetry.
The leading analytic correction to compressibility in this case is proportional to $B^2$.~\cite{Spivak12}
However, the nonanalytic $\sim |B|$ correction still exists. The same is true in the special case of Weyl semimetal in Y$_2$Ir$_2$O$_7$, where, while 
time reversal symmetry is broken, the macroscopic magnetization is zero due to the ``all-in/all-out" magnetic order.~\cite{Wan11} 
The presence of such a nonanalytic $\sim |B|$ correction to compressibility
and to (square of) the plasmon frequency may thus be regarded as a {\em universal property of Weyl metals}, independent of the specific
realization.
This effect is especially striking when the Fermi level coincides with the Weyl nodes, i.e. $\epsilon_F = 0$ in our model.
In this case, in the absence of magnetic field the Weyl semimetal is incompressible, since the density of states vanishes.
However, when an external field is applied, there appears a finite compressibility, proportional to $|B|$ and the corresponding
plasmon collective mode with frequency $\omega \sim \sqrt{|B|}$.

\begin{acknowledgments}
Financial support was provided by NSERC of Canada and a University of Waterloo start up grant.
AAB and DAP acknowledge the hospitality of the Aspen Center for Physics, funded by NSF grant PHY-1066293, where this work was initiated. 
\end{acknowledgments}

\end{document}